# Big Data in IoT Systems

**Fayeem Aziz, Stephan K. Chalup and James Juniper**

# Big Data in IoT Systems

**Fayeem Aziz, Stephan K. Chalup and James Juniper**
Interdisciplinary Machine Learning Research Group
The University of Newcastle, Callaghan, NSW 2308, Australia
stephan.chalup@newcastle.edu.au

## 2.1 Introduction

Big Data in IoT is a large and fast-developing area where many different methods and techniques can play a role. Due to rapid progress in Machine Learning and new hardware developments, a dynamic turnaround of methods and technologies can be observed. This overview therefore tries to be broad and high-level without claiming to be comprehensive. Its approach towards Big Data and IoT is predicated on a distinction between the digital economy and the characteristics of what Robin Milner has described as the Ubiquitous Computing System (UCS) (Milner, 2009).

### 2.1.1 The Digital Economy and Ubiquitous Computing Systems

The characteristics of a UCS are that (i) it will continually make decisions hitherto made by us; (ii) it will be vast, maybe 100 times today's systems;

---

(iii) it must continually adapt online to new requirements; and (iv) individual UCSs will interact with one another (Milner, 2009). Milner (2009) defines the UCS as a system with a population of interactive agents that manage some aspect of our environment. In turn, these software agents move and interact, not only in physical space, but also in virtual space. They include data structures, messages and a structured hierarchy of software modules. Milner's formal vision is of a tower of process languages that can explain ubiquitous computing at different levels of abstraction. Generic features of the contemporary UCS include concurrency, interaction, and decentralized control.

The notion of the digital economy has been clearly articulated in defining Germany's Industry 4.0 program. In Industry 4.0 manufacturing management and software industry converge into a joint concept that combines IT, Big Data analytics, and production on a global scale. While industrial manufacturing machines communicate within IoT, human technicians should have the ability to check on production and process quality locally at a production floor site and eventually make real-time decisions based on complex analytics provided by the global industry 4.0 data analytics components. Cloud-based smart watch software (Gottwalles, 2016) allows for this and integrates local technician into the Industry 4.0 IoT.

Global digital factory software systems for Industry 4.0 optimization have become a central control tool developed by leading manufacturers and software developers such as Siemens, Bosch, Kuka, SAP and Fraunhofer IPA. Product life cycle management provides information management systems that integrate data, processes, business systems, and employees in a digital factory. While real-time monitoring, analysis, and traceability constitute one set of aspects where Big Data techniques and machine learning come into play, another aspect is prediction and modelling. Associated techniques could come into play before starting a new and large industry component or as a sophisticated tool to predict service requirements. Again, large software systems with Big Data analytics modules are required to run virtual simulations of complex production processes, logistics, distribution, financial risk, equipment health and human aspects, etc.

Individual manufacturers can provide detailed digital models of mechanical devices and production robots. Fog computing or real-time edge computing is a software layer above machine/robot control software but below cloud and a hierarchy of process control systems, manufacturing execution systems, and enterprise resource planning. Fog computing includes communications, analysis, control, and orchestration of machine control of endpoints on the industrial floor and also connects to the management of fleets and warehouses. Big Data analytics techniques can be employed to analyze production data at the top level by connecting enterprise resource planning and cloud data. Machine learning techniques for big data are very general and can potentially be used at each level mentioned in order to increase not only efficiency, but also aspects of security and safety. Industry 4.0 software systems integrate all these various aspects and can also help manufacturers to adapt their systems to new energy regulations, national policies and keep global control over local robot life-cycles.

The concept of Industry 4.0 is still new, complex and fascinating. Future developments may consider the integration and release of General AI based decision modules that are based on their wide connectivity to all levels and sections of the system. Their fast processing capacities (using Big Data techniques) could exhibit superhuman abilities in real-time decision making. These autonomous Industry 4.0 decision modules would outperform humans just as high-speed trading robots already outperform human traders in the stock market. These modules would lead to better productivity of the digital factory and also to more safety and efficiency. They may become a necessary component to achieve an efficient working global system in times of demographic change, resource shortages and environmental challenges.

To provide an overview of the various domains of digital communication associated with Big Data and the IoT Systems, the following diagram (Fig. 2.1) depicts two areas of intersection. The pale slanting lines mark out the intersection between "Digital Communication" and "The Internet," while the dark slanting lines identify the domain of intersection between "Big Data," the "Internet of Things" (IoT), and "Machine-to-Machine" Communication. Other chapters in this text deal at length with digital communication.

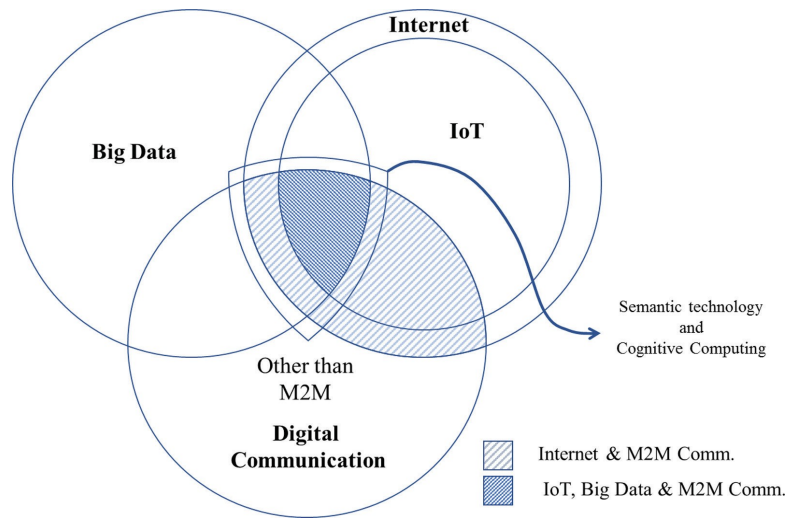

Figure 2.1 Big Data in IoT systems and machine-to-machine communications.

Therefore, this chapter will primarily focus on Semantic Technologies (which extract information from the World-Wide Web using diagrammatic reasoning for purposes of business intelligence) and cognitive computing systems (which represent the dominant form of artificial intelligence involving deep machine learning that integrates sensing or perception with action). These computer- based technologies are probably best thought of as applying to the Internet as a whole, as indicated by the label on the trisected field in the middle of the diagram. They could even apply to digital communication as a whole, given that business intelligence can obviously be transferred across divisional, spatial, and functional boundaries within any given organization.

### 2.1.2 The Internet of Things

The Internet of Things (IoT) is one of the most rapidly emerging platforms for the digital economy (Juniper, 2018). It is a web- based network, which connects smart devices for communication, data transfer, monetary exchange, and decision-making. Both the number of communication channels and the volume of data transmitted are increasing exponentially along with the number of devices that are connected to this network. According to Forbes (May 2014),

> *By 2020 there will be over 26 billion connected devices . . . That's a lot of connections (some even estimate this number to be much higher, over 100 billion). The IoT is a giant network of connected "things" (which also includes people). The relationship will be between people-people, people-things, and things-things. (Morgan, 2014)*

Many developed countries are applying or planning to apply IoT to smart homes and cities. For example, Japan provides dedicated broadband access for "things-to-things" communication, while South Korea is building smart home control systems that can be accessed remotely. The IoT European Research Cluster (IERC) has proposed a number of IoT projects and created an international IoT forum to develop a joint strategic and technical vision for the use of IoT in Europe (Santucci, 2010). China is planning to invest $166 billion in IoT industries by 2020 (Voigt, 2012). Figure 2.2 shows publication-based research trends involving IoT. The growing IoT produces a huge amount of data that will need to be processed and analyzed.

For the processing and analysis of very large data sets—Big Data—a new research area and associated collection of methods and techniques have emerged in recent years. Although there is no clear definition for Big Data, a commonly quoted characterization are the "3V's": volume, variety, and velocity (Laney, 2001; Zaslavsky et al., 2012).

- Volume: There is more data than ever before. Its volume continues to grow faster than we can develop appropriate tools to process it.
- Variety: There are many different and often incompatible types of data such as text data, sensor data, audio and video recordings, graphs, financial, and health data.
- Velocity: Data can be streaming, that is, it is arriving continuously in real time and we are interested in obtaining useful information from it instantly. The ability to process depends not only on physical bandwidth and protocols but also on suitable architectural solutions and fast algorithms.

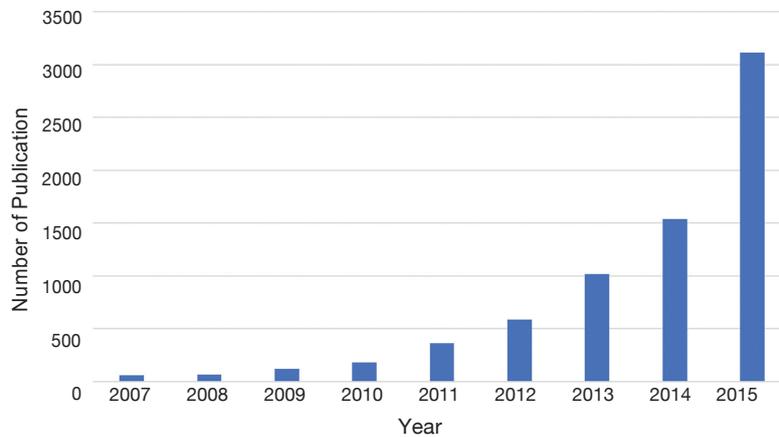

Figure 2.2 Current research trend in IoT based on number of publications (Source: Web of knowledge).

More recently, at least two more Vs have been added to the list of Big Data criteria (Fan & Bifet, 2013; Tsai et al., 2015):

- Variability: Data has variation in structure and interpretation depending on the applications.
- Value: Data has an effective business value that gives organisations a competitive advantage. This is due to the ability of making decisions based on extensive data analysis that was previously considered beyond reach.

IoT data satisfies the criteria of the "V-defined" big-data category. It has been predicted by several authors that the large number of connected objects in IoT will generate an enormous amount of data (Botta et al., 2016; Dobre & Xhafa, 2014; B. Zhang et al., 2015). The IoT-generated data are variable in terms of structure, often arrive at real-time, and might be of uncertain provenance. These large amounts of data require classification, processing, analysis, and decision-making engines for commercially viable usage. It will be necessary to develop techniques that convert this raw data into usable knowledge. For example, in the medical area, raw streams of sensor values must be converted into semantically meaningful activities such as eating, poor respiration, or exhibiting signs of depression performed by or about a person (Stankovic, 2014).

## 2.2 Theoretical Approaches to UCS

The diversity and rapid, if not chaotic, development of UCS make it hard to capture it within formal engineering frameworks. We discuss additional theoretical viewpoints that may help with the daunting task of maintaining an overview on further developments of Big Data in IoT.

### 2.2.1 Category Theory

Category theory—a branch of pure mathematics that weaves together formal representations of structures and dynamic transitions between structures that can be found in algebra, geometry, topology, computation, and the natural sciences—is often portrayed as an advance over earlier foundational approaches to mathematics that were grounded in Set Theory (Bell, 1988; Krömer, 2007; Marquis, 2009; Rodin, 2012).

Category theory provides Big Data and IoT with a variety of computational frameworks including the co-algebraic representation of automatons and transition systems, domain theory, the geometry of interaction, along with specific sites such as elementary topos. Categorical logic links inferential procedures and resource- using logics with functional programming, while string diagrams can represent everything from graphical linear algebra to signal flow graphs and functional relationships in topological quantum field theory.

### 2.2.2 Processes Algebras

Another approach to the formal modelling of concurrent and communicative systems is process algebra which, over time, has drawn on various calculi of interactive, sequential, concurrent, and communicative systems. In his history of process algebra, Baeten (2005) observes that on

> *comparing the three most well-known process algebras CCS (Calculus of Communicative Systems), CSP (Calculus of Sequential Processes) and ACP (Algebra of Communicative Processes), we can say there is a considerable amount of work and applications realized in all three of them. Historically, CCS was the first with a complete theory. Different from the other two, CSP has a least distinguishing equational theory. More than the other two, ACP emphasizes the algebraic aspect: there is an equational theory with a range of semantical models. Also, ACP has a more general communication scheme: in CCS, communication is combined with abstraction, in CSP, communication is combined with restriction.*

Contemporary approaches to business process modeling are typically based on stochastic versions of Milner's pi calculus or stochastic Petri nets. Towards the end of a remarkably productive life, Milner attempted to formally merge both these calculi together using the framework of bigraphs. Many careers in computational research have been grounded in efforts to build bridges between process algebras, functional programming, linear logic, and monoidal categories.

## 2.3  Core Digital Technologies for UCS

### 2.3.1  Semantic Technologies

Semantic technologies provide users with integrated access to data by applying search and navigation techniques that are tuned to the computational ontologies of relevance to the organization. To this end, it draws on the WC3 standards for the World Wide Web, in accordance with which the Resource Description Framework is formally conceived as a "giant global graph" (Connolly, 2010; Grau et al., 2012). Diagrammatic reasoning procedures and visual analytic processes are applied to support business intelligence.

A recent example is the CUBIST Project. The CUBIST project largely drew on Peirce's Existential Graphs (Dau & Andrews, 2014). It brings together a consortium of Technological Partners that includes: SAP—Germany (Coordinator and technological partner); Ontotext—Bulgaria (providing expertise in Semantic Technologies); Sheffield Hallam University—UK (providing expertise in Formal Concept Analysis); Centrale Recherche S.A.—France (providing expertise in FCA and Visual Analytics) and Case Partners that include Heriot-Watt University—UK (providing expertise in the analysis of gene expressions in mouse embryos); Space Applications Services— Belgium (providing expertise in the analysis of logfiles of technical equipment in space along with space system engineering, specification, operations engineering, training and software development) and Innovantage—UK (providing expertise in the analysis of the online recruitment activities of UK companies).

The core objective of the project is to investigate "how current semantic technologies can be applied in enterprise environments to semantically integrate information from heterogeneous data sources and provide unified information access to end users."

Under the architecture of the CUBIST Prototype, there are different means of access to information, including through semantic searching based on the domain ontologies specific to each of the three case studies: "smart" query generation taking these computational ontologies into account, where the types and object properties form a "query graph" that can actually contain more types than those selected, with their associated datatype properties being used for the filtering and characterization of formal attributes; more explorative search techniques; conceptual scaling as described above; and visual analytics.

To this end, it draws on the Resource Description Framework of the WWW. In this context, it pursues "semantic integration," which essentially means transforming the information into a graph model of typed nodes (e.g., for products, companies) and typed edges (e.g., for the relationship "company-produces-product"), then performing formal concept analysis (FCA) on the transformed information. In this way, it aims to provide unified access by letting users search, explore, visualize, and augment the information as if it was from one single integrated system.

### 2.3.2 Cognitive Computing and Deep Learning

With the availability of large amounts of data and the ability to process it efficiently using GPU technologies, deep learning led to surprising advances in machine learning and pattern recognition applications. A key breakthrough example was the outstanding performance of a deep convolutionary neural net with 650,000 neurons in the ImageNet Large-Scale Visual Recognition Challenge (Krizhevsky et al., 2012). A subset of 1.2 million labeled (256x256) images was used for training, while the full ImageNet dataset consists of 15 million labeled high-resolution images in 22,000 categories. More recent deep networks can be much larger and use even more data. It can be said that big data enables deep learning and we can expect exiting applications of deep learning technology to the large amounts of data produced by IoT in the near future.

Bengio et al. (2013) consider recent advances in unsupervised learning and deep learning, canvassing three major approaches that have been adopted towards deep networks: (i) advances in probabilistic models; (ii) directed learning (using sparse coding algorithms) and undirected learning (i.e., Boltzmann machines); and (iii) auto-encoders (reconstruction-based algorithms) and manifold learning (geometrically-based). Bengio et al. warn that, at present, successful outcomes still depend heavily on taking advantage of "human ingenuity and prior knowledge" to compensate for the weakness of current learning algorithms. Tohmé and Crespo (2013) also warn that,

> *Computational intelligence only provides rough approximations to the task of theory or model building Systems like BACON (in any of its numerous incarnations) despite their claimed successes (and) are only able to provide phenomenological laws (Simon 1984). That is, they are unable to do more than yield generalizations that involve only observable variables and constants. No deeper explanations can be expected to ensue from their use.*

Bengio et al. (2013) concede that "it would be highly desirable to make learning algorithms less dependent on feature engineering, so that novel applications could be constructed faster, and more importantly, to make progress towards Artificial Intelligence (AI)." In this light they note a string of recent successes in speech recognition, signal processing, object recognition, natural language processing, and transfer learning. However, it is not yet clear whether these advances are always adequate to the task.

Similar concerns on the limitations of current deep structured learning, hierarchical learning, and deep machine learning are expressed by Michael Jordan, Chair of the National Academy's "Frontiers in Massive Data Analysis" Committee. After describing deep learning as an attempt to model high level abstractions in data via multiple processing layers, each composed of multiple linear and non-linear transformations, and using efficient algorithms for un-/semi-supervised feature learning and hierarchical feature extraction, with some approaches inspired by advances in neuroscience, Jordan notes the need for a certain hard-headed realism in cautioning that

> *The overeager adoption of big data is likely to result in catastrophes of analysis comparable to a national epidemic of collapsing bridges. Hardware designers creating chips based on the human brain are engaged in a faith-based undertaking likely to prove a fool's errand. Despite recent claims to the contrary, we are no further along with computer vision than we were with physics when Isaac Newton sat under his apple tree.*

Bengio et al. (2013) explain the structural conditions that are necessary to arrive at a successful representation, namely:
- smoothness of function f to be learned;
- in multi-factor explanations, learning about one factor generalizes to learning about others;
- the existence of an abstractive hierarchy of representations; in semi-supervised learning representations, they observe that what is useful for P(X) tends to be useful for P(Y|X);
- for learning tasks, P(Y|X, task)'s ought to be explained by factors shared with other tasks;
- for manifolds, probability mass should concentrate near regions that have a much smaller dimensionality than the original space where the data lives;
- in natural clustering, the $P(X|Y_i)$'s for different i tend to be well separated;
- consecutive or contiguous observations tend to be associated with similar values for relevant categorical concepts;

- sparsity is achieved such that, for any given observation x, only a small fraction of the possible factors are relevant;
- and simplicity of factor dependencies are obtained so that, in good high-level representations, the factors are related to each other through simple, typically linear dependencies.

Of course, the very factors that help to explain success also help to identify the conditions for failure. A burgeoning literature (Goertzel, 2015; Nguyen et al., 2015; Szegedy et al., 2013) has identified the "hallucinatory" capacity of deep learning networks to assign high levels of statistical significance to "recognized" features that are not even present in the data. This is brought home most graphically in some of Hern's (2015) visual examples of image misrecognition.

Bengio et al. (2013) note that techniques for greedy, layerwise, unsupervised, pre-training were a significant technical breakthrough, enabling deep learning systems:
- to learn a hierarchy of features one level at a time, using unsupervised feature studies to learn a new transformation at each level.
- to be composed with the previously learned transformations (i.e., each iteration of unsupervised feature learning adds one layer of weights to a deep neural network)
- so that the set of layers can be combined to initialize a deep supervised predictor, such as a neural network classifier, or a deep generative model such as a deep Boltzmann machine.

A more sanguine appraisal of these developments in deep learning—one that recognizes the pertinence of the critique of artificial intelligence mounted by Dreyfus (2005)—would help to explain why deep learning will continue to rely heavily on human intervention into the future (Dreyfus, 2005). It also explains why researchers in the field of machine learning have set themselves fairly modest objectives. For example, Bottou's (2014) plausible definition of "reasoning" entails "algebraically manipulating previously acquired knowledge in order to answer a new question." On this view, machine reasoning can be implemented by algebraically enriching "the set of manipulations applicable to training systems" to "build reasoning capabilities from the ground up." The example he describes is one involving an optical character recognition system constructed "by first training a character segmenter, an isolated character recognizer, and a language model, using appropriate la- belled training sets," then "adequately concatenating these modules and fine tuning the resulting system can be viewed as an algebraic operation in a space of models". He observes that the resulting model can answer a new question, namely, "converting the image of a text page into a computer readable text."

Along similar lines, a team working on AI for Facebook (Lopez-Paz et al., 2016) on the task of discovering causal signals in images have built a classifier that "achieves state-of-the-art performance on finding the causal direction between pairs of random variables, when given samples from their joint distribution." This "causal direction finder" is then deployed "to effectively distinguish between features of objects and features of their contexts in collections of static images."

In the context of IoT the question could be asked if deep learning–based systems would be able to have a "superhuman" look at the extracted data and could come to any useful conclusions or regulatory actions. Systems like AlphaGo Zero (Silver et al., 2017) that involve deep reinforcement learning could be employed in a multi-agent setting and optimize communication of system components or group behavior.

## 2.4 Big Data and Its Sources

Unlike the conventional Internet with the standard specification, currently IoT does not have a defined system architecture (Huansheng Ning & Hu, 2012). In the last five years, different types of structures have been proposed for IoT system architecture including layer-based models, dimension-based models, application domain structures and social domain structures (Luigi Atzori et al., 2010; Huansheng Ning & Hu, 2012). Layer-based models are the most commonly used structures in the literature of IoT and layers in the structure are typically sensor layers, network layers, service layers and interface layers (Atzori et al., 2011; Bermudez-Edo et al., 2016; Lu & Neng, 2010; Miao et al., 2010; Ning & Wang, 2011; Xu et al., 2014). IoT system architecture comprises radio frequency identification, wireless sensor networks, middleware software, cloud computing, and IoT application software as depicted in Fig. 2.3 (Lee & Lee, 2015). These aspects of Big Data and IoT technologies are described briefly in the following subsections.

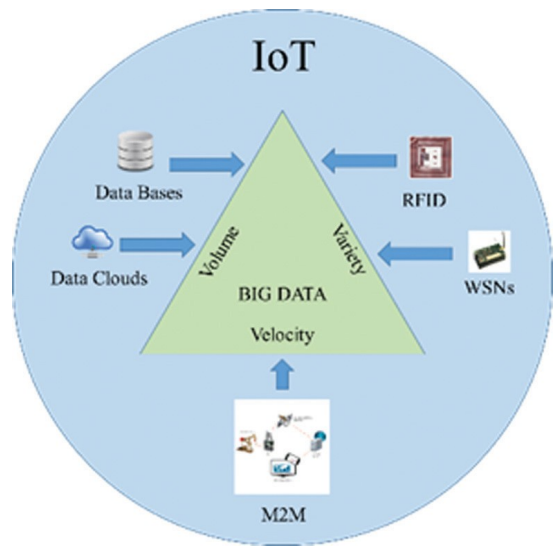

Figure 2.3 Major sources of Big Data gathering in the IoT.

### 2.4.1 Radio Frequency Identification

Radio frequency identification (RFID) allows automatic identification and data capture using radio waves, a tag, and a reader. The tag can store more data than traditional barcodes. The tag contains data in the form of a global RFID-based item identification system developed by the Auto-ID Center (Khattab et al., 2017).

Database entries for tags can have an effectively unlimited size. Therefore, the size of the database of a tag and its associated object can be enormous (Juels, 2006). For instance, in modern manufacturing plants processes use RFID tagged resources. These resources generate a large amount of logistic data while they move through the production process (Russom, 2011). The analysis of this enormous amount of data can reveal significant information and suggestions in improving logistics planning and layout of distribution (Zhong et al., 2015).

### 2.4.2 Wireless Sensor Networks

Wireless sensor networks (WSNs) consist of spatially distributed autonomous sensor-equipped devices to monitor physical or environmental conditions and can cooperate with RFID systems to better track the status of things such as their location, temperature, and movements (Luigi Atzori et al., 2010). Recent technological advances in low-power integrated circuits and wireless communications have made available efficient, low-cost, low-power miniature devices for use in WSN applications (Gubbi et al., 2013). WSN provide a virtual layer through which the digital systems can access information of the physical world. Therefore, WSNs have become one of the most important elements in IoT (Gimenez et al., 2014).

WSNs gather large quantities of real-time data by various types of sensors such as proximity sensors, humidity sensors, thermal sensors, magnetic, position, and flow sensors. WSNs have become an important technology to support the gathering of big data in indoor environments where they can collect information, for instance, on temperature, humidity, equipment working conditions, health inputs, and electricity consumption (Ding et al., 2016; Rani et al., 2017). Big Data mining and analysis algorithms are specialized on processing and managing these immense volumes of data for various operations (Gimeez et al., 2014; Rani et al., 2017). For instance, car-manufacturing companies are mounting various sensors on their manufactured cars for surveillance of the product. Data collected from the sensors are being stored in a web server and middleware software analyzes the data to assess performance and in detecting defects of their manufactured cars and their parts.

### 2.4.3 Machine-to-Machine Communications

Machine-to-machine communications (M2M) represent a future where billions of everyday objects and information from the surrounding environment are connected and managed through a range of devices, communication networks, and cloud-based servers (Wu et al., 2011).

Mathew et al. (2011) describe a simple architecture for the Web of Things (WoT), where all objects are connected to a knowledge-based server. The WoT is the simplest early version of IoT. Bell labs presented a prototype implementation of the WoT with four layers: physical objects, a WoT browser, application logics, and virtual objects (Christophe et al., 2011).

As M2M sensors have limited storage and energy capacity, their networks require transmission of a large amount of real-time data. This data-transmission needs to address the issues of efficiency, security, and safety (Suciu et al., 2016). Various proposals have been put forward to solve these issues in M2M communication. For example, knowledge management–integrated big data channels have been proposed (Sumbal et al., 2017).

### 2.4.4 Cloud Computing

Cloud computing is a model for on-demand access to a shared pool of configurable resources (e.g., computers, networks, servers, storage, applications, services, software) that can provide Infrastructure as a Service, Software as a Service, Platform as a Service or Storage as a Service (Suciu et al., 2016). Accordingly, IoT applications require massive data storage, a high speed to enable real-time decision- making, and high-speed broadband networks to stream data (Lee & Lee, 2015; Gubbi et al., 2013).

## 2.5 Big Data in IoT Application Areas

### 2.5.1 Healthcare Systems

IoT is providing new opportunities for the improvement of healthcare systems by connecting medical equipment, objects, and people (Zhibo et al., 2013). Technological developments associated with wireless sensors are making IoT-based healthcare services accessible even over long physical distances.

Web-based healthcare or eHealth services are sometimes cheaper and more comfortable than conventional face-to-face consulting (Hossain & Muhammad, 2016; Sharma & Kaur, 2017). Moreover, IoT's ubiquitous identification, sensing and communication capabilities means that all entities within the healthcare system (people, equipment, medicine, etc.) can be continuously tracked and monitored (Alemdar & Ersoy, 2010; Mohammed et al., 2014).

Cloud computing, Big Data and IoT and developing ICT artifacts can be combined in shaping the next generation of eHealth systems (Suciu et al., 2015). Processing of large amounts of heterogeneous medical data, which are collected from WSNs or M2M networks, supports a movement away from hypothesis-driven research to- wards more data-driven research. Big data search methods can find patterns in data drawn from the monitoring and treatment of particular health conditions.

A detailed framework for health care systems based on the integration of IoT and cloud computing has recently been described and evaluated (Abawajy & Hassan, 2017). In accordance with this approach, lightweight wireless sensors are installed in everyday objects such as clothes and shoes, to observe each patient's physiological parameters such as blood sugar levels, blood glucose, capnography, pulse, and ECG. The data that is collected is then stored in personalized accounts on a central server. This server provides a link between the IoT subsystem and the cloud infrastructure. In the cloud, various data analysis programs have been installed to process the information for clinical observation and notify emergency contacts if and when an alarm is triggered. Other programs such as analytics engines extract features and classifies the data to assist healthcare professionals in providing proper medical care (Abawajy & Hassan, 2017).

In the field of clinical management, the main benefits pro-vided by these interacting systems include (i) improved decision-making about effective treatment, (ii) early detection of errors in treatment, (iii) improved assessment of the performance of medical professionals, (iv) the development of new segmentation and predictive models that incorporate unit record data on patient profiles, (v) automation of the payment system and cost control, and (vi) the transmission of information to the right people at the appropriate time.

Diagnosis will also be improved because each health center can access the requisite patient information regardless of where the tests are conducted. Moreover, test data can be stored in real time, allowing decisions to be made from the instant that a test has been completed. By dramatically reducing the storage and processing time, feasible Big Data techniques can also support research activity. NOSQL technologies that are focused on the patient will also allow monitoring and storage of data collected from both inside and outside the home, with early warnings on changes in health status and alarm systems identifying the need for preventive action leading to cost savings by reducing the number of emergency visits and the length of resulting hospital stays.

### 2.5.2 Food Supply Chains

Existing food supply chains (FSC) are very complex and widely dispersed processes that involve a large number of stakeholders. This complexity has created problems for the management of operational efficiency, quality, and public food safety. IoT technologies offer promising potential to address the traceability, visibility, and controllability of these challenges in FSC (Gia et al., 2015; Xu et al., 2014) especially through the use of barcode technologies and wireless tracking systems such as GPS and RFID at each stage in the process of agricultural production, processing, storage, distribution, and consumption. A typical IoT solution for FSC comprises three parts:
- field devices such as WSNs nodes, RFID readers/tags, user interface terminals, etc.;
- backbone systems such as databases, servers, and many kinds of terminals connected by distributed computer networks, etc.; and
- communication infrastructure such as WLAN, cellular, satellite, power line, Ethernet, etc. (Xu et al., 2014).

### 2.5.3 Smart Environment Domain

#### 2.5.3.1 Smart power system

With advances in IoT technology, smart systems, and Big Data analytics, cities are evolving to become "smarter" (Stankovic, 2014). For example, patterns of the power usage households can be monitored and analysed across different time-periods to manage the cost of power (Rathore et al., 2017). According to recent research, smart grid technology is one feasible solution helping to overcome the limitations of traditional power grid systems (Iyer & Agrawal, 2010; Parikh et al., 2010; Stojkoska & Trivodaliev, 2017).

#### 2.5.3.2 Smart home

Stojkoska and Trivodaliev have outlined and proposed a generalized framework for an IoT-based smart home (Stojkoska & Trivodaliev, 2017). Their framework connects the home, utilities and third-party application providers through to a cloud network, with sensors attached to the smart grid system gathering data from smart home appliances. As most utilities apply time-of-use charges ("Understand energy prices & rates," 2017), third-party application providers can reduce utility costs by combining appliances such as battery chargers with refrigerators, and ovens that can be controlled over the web (Buckl et al., 2009). This also applies to renewable energy sources with web-based meters calculating how much power the home will require from the grid. The smart home framework is shown in Fig. 2.4.

#### 2.5.3.3 Smart environment control

Many manufacturing enterprises have strict requirements on equipment working conditions and environment conditions for high-quality products, especially in chip fabrication plants, pharmaceutical factories, and food factories (Ding et al., 2016). In the product manufacturing process, data on working condition variables and environmental conditions need to be gathered, stored and analyzed in real time to identify risks and abnormalities. Predictive and remotely controlled manufacturing systems also consist of integrated platforms for predictive analytics and visualization of data derived from analytic engines (Lee et al., 2013).

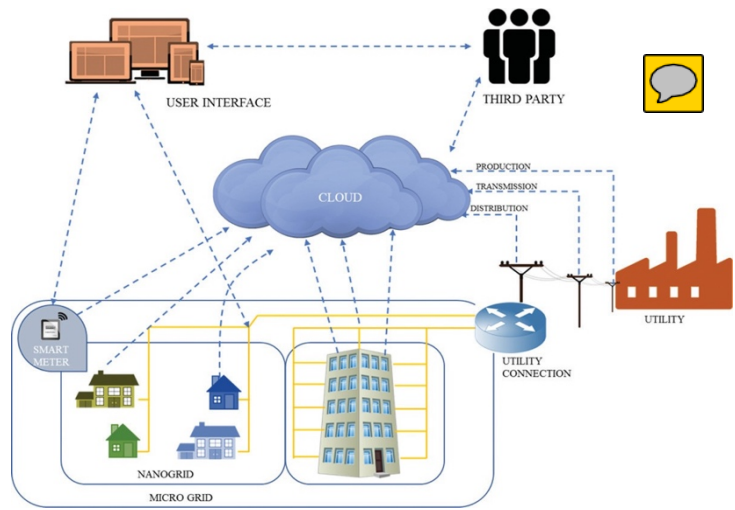

Figure 2.4 Multi level IoT framework for smart home (Stojkoska & Trivodaliev, 2017).

A WSN was developed for vegetable greenhouse monitoring and a control system for agriculture (Srbinovska et al., 2015). This system helps farmers to increase crop production and crop quality by remotely controlling different components of the greenhouse such as drip irrigation and fan facilities. In (Stojkoska et al., 2014), the authors present a framework for temperature regulation inside commercial and administrative buildings, which focuses on the design and implementation of specific sensory network topologies and nodes within the system.

### 2.5.3.4  Safety and surveillance

Smart environments also help in improving safety aspects of the automation process in industrial plants through a massive deployment of RFID tags associated with each of the production parts (Spiess et al., 2009). Furthermore, for safety, surveillance, process monitoring and security purposes, the sensing of gasphase particles has been important in certain processes. While portable instruments can detect a diverse range of gas particles, connected multivariable sensors are a relatively new but effective method in issuing warnings on potential disasters, both in industrial environments and at home (Potyrailo, 2016).

### 2.5.3.5 Smart city

Rapid growth of city populations due to urbanization has resulted in a steady increase in connectivity which in turn has generated a massive and heterogeneous amount of data. Increasingly, Big Data analytics is providing a better understanding of urban activities to support both current management and future planning and development. Rathore et al. (2017) envision a "Super City" that is both smarter and safer than current conceptions of a smart city. In Super City, residents and workers are supported in their actions anytime, anywhere, for any purpose, including ensuring that they are more secure and safe from theft, robbery, assaults and other crimes as well as from external environmental threats such as pollution. Rathore et al.'s Super City planning includes an IoT with a four-tiered model for Big Data analytics that comprises data generation and collections, data communication, data administration and processing, and data interpretation (Rathore et al., 2017).

### 2.5.4 Safer Mining Production

Mine safety is a major concern for many countries due to the high risk working conditions in underground mines. In this context, IoT technology can be used to detect signs of a potential mine disaster due to flooding, fires, gas explosions, dust explosions, cave, coal and gas outbursts, leakage of toxic gases, and various other risk factors (Qiuping et al., 2011). Once again, RFID, WiFi, and other wireless communications technologies and devices are deployed to enable effective communication between surface and underground, to track the location of underground workers and analyze critical safety data collected from chemical and biological sensors.

### 2.5.5 Transportation and Logistics

The transportation and logistics industry is undergoing enormous technological changes occasioned by the introduction of tracking and tracing technologies. RFID and NFC technologies can be deployed for real-time monitoring of almost every link in the supply chain, ranging from commodity design, raw material purchasing, production, transportation and storage, through to distribution, sale of semi-products and products, returns processing, and after-sales service (Luigi Atzori et al., 2010). In particular, instant tracking of package delivery is reducing transfer time across different layers of the transport system, with courier services providing immediate tracking through mobile phone apps.

WSNs are used in cold chain logistics that employ thermal and refrigerated pack-aging methods to transport temperature-sensitive products (Hsueh & Chang, 2010). Zhang et al. (2012) have designed an intelligent sensing system to monitor temperature and humidity inside refrigerator trucks by using RFID tags, sensors, and wireless communication technology.

WSNs are also used for maintenance and tracking systems. For example, General Electric deploys sensors for the preventive maintenance of its jet engines, turbines, and wind farms. Likewise, American Airlines uses sensors capable of capturing 30 terabytes of data per flight for this purpose. Car manufacturers are mounting infrared, heat pressure, and other sensors to monitor the health of a car, while GPS devices are providing position information to determine traffic density and navigation assistance (Qin et al., 2013).

The idea of driverless cars is central to planning the future of our transportation. Driverless cars are connected to the network using WSN technologies to provide data from their sensors and to receive feedback after data analysis. The cars can access information from a database of maps and satellite information for GPS localization and global traffic and transport demand optimization. Critical for safety is communication between cars that navigate in close proximity, with crucial data from vision sensors processed on-board and in real-time using compact, high-performance computing devices such as GPU cards.

### 2.5.6 Firefighting

IoT has been used in the firefighting safety field to detect potential fires and provide early warnings of possible fire-related disasters. RFID tags and barcodes on firefighting items, mobile RFID readers, intelligent video cameras, sensor networks, and wireless communication networks are used to build a database for nationwide firefighting services (Zhang & Yu, 2013).

## 2.6 Challenges of Big Data in IoT Systems

Big Data usually requires massive storage, huge processing power and can cause high latency (Lee et al., 2013). These challenges are demanding Big Data specific processing and computation layers in the IoT chain (Bessis & Dobre, 2014; Samie et al., 2016). In addition, a closer inspection of IoT revealed issues not only with scalability, latency, bandwidth, but also with privacy, security, availability, and durability control (Fan & Bifet, 2013).

The challenges of handling Big Data are critical since the overall performance is directly proportional to the properties of the data management service. Analyzing or mining massive amounts of data generated from both IoT applications and existing IT systems to derive valuable information requires strong Big Data analytics skills, which could be challenging for many end-users in their application and interpretation (Dobre & Xhafa, 2014). Integrating IoT devices with external resources such as existing software systems and web services requires the development of various middleware solutions as applications can vary substantially with industries (Gama et al., 2012; Roalter et al., 2010).

### 2.6.1 Big Data Mining

Extracting values from Big Data with data mining methodologies using cloud computing now typically requires the following (Rashid et al., 2017; Triguero et al., 2015; Yang et al., 2015; Q. Zhang et al., 2015):
- detecting and mining outliers and hidden patterns from Big Data with high velocity and volume
- mine geospatial and topological networks and relationships from the data of IoT

- developing holistic research directed at the distribution of traditional data mining algorithms and tools to cloud computing nodes and centers for Big Data mining
- developing a new class of scalable mining methods that embrace the storage and processing capacity of cloud
- platforms addressing spatiotemporal data mining challenges in Big Data by examining how existing spatial mining techniques succeed or fail for Big Data
- providing new mining algorithms, tools, and software as services in the hybrid cloud service systems

### 2.6.2 Proposed Big Data Management and Analysis Techniques

The current approaches of data analysis demand benchmarking the databases including graph databases, key-value stores, time- series and others (Copie et al., 2013). Heterogeneous addressing systems of objects such as wireless sensors are creating complex data retrieval processes. End users are demanding a homogenous naming and addressing convention for objects so that they can retrieve and analyze data regardless of the platforms or operating system (Liu et al., 2014). IPv4, IPv6, and Domain Name Service (DNS) are usually considered as the candidate standard for naming and addressing; however, due to the lack of communication and processing capabilities of many small and cheap devices (like RFID tags) it is quite challenging to connect everything with an IP.

SQL-based relational databases provide centralized control of data, redundancy control and elimination of inconsistencies. The complexity and variability of Big Data require alternative models of databases. Primarily motivated by the issue of system scalability, a new generation of databases known as NoSQL is gaining strength and space in information systems (Vera et al., 2015). NoSQL are database solutions that do not provide an SQL interface.

Cloud computing provides fundamental support to address the challenges with shared computing resources including computing, storage, networking and analytical software; the application of these resources has fostered impressive Big Data advancements (Yang et al., 2017). The Mobile Cloud (MC) is emerging as one of the most important branches of cloud computing and is expected to expand mobile ecosystems. MC is a combination of cloud computing, mobile computing, and wireless networks designed to bring rich computational resources to mobile users and network operators as well as cloud computing providers (Chen, 2015; Han et al., 2015; Hasan et al., 2015; Nastic et al., 2015). Mobile devices can share the virtually unlimited capabilities and resources of the MC to compensate for its storage, processing, and energy constraints. Therefore, researchers have predicted that MC would be one of the complementary parts of IoT to provide a solution as a big database (Bonomi et al., 2012; Botta et al., 2016; Singh et al., 2014). As a result, an integration of MC and IoT is emerging in current research and is called the MCIoT paradigm (Kim, 2015).

A four-tiered Big Data analytical engine is designed for the large amounts of data generated by the IoT of smart cities (Paul, 2016; Rathore et al., 2017). The Hadoop platform performs extraordinarily when used in the context of analyzing larger datasets (Rathore et al., 2017). A Hadoop distributed file system in IoT-oriented data storage frameworks allows efficient storing and managing of Big Data (Jiang et al., 2014). A model that combines Hadoop architecture with IoT and Big Data concepts was found to be very effective in increasing productivity and performance in evolutionary manufacturing planning which is an example of Industry 4.0 (Vijaykumar et al., 2015).

## 2.7  New Product Development and UCS

The growth of the UCS has transformed the process of New Product Development. In their comprehensive history of Iterative and Incremental Development, Larman & Basili (2003) trace SCRUM and "Agile" techniques of New Product Development back to software engineering projects in the 1950s.

However, despite decades of criticism from software engineers and contractors, the DoD only changed their waterfall-based standards at the end of 1987 to allow for iterative and incremental development on the basis of recommendations made in a report published in October of that year by the Defense Science Board Task Force on Military Software, chaired by Frederick Brooks (Larman & Basili, 2003). In his representative critique of the water fall model (which is depicted schematically below), Goguen (1994) insists that there should be no presumption of an orderly progression from one stage to the next, suggesting instead that it is more of a zigzagging backwards and forwards, with phases constantly overlapping in circumstances where managers have great difficulty in assigning actions and events to specific phases. Moreover, he observes that the requirements engineering phase of software development is critical as it is the costliest, the most error prone, the most exposed to uncertainty, and, therefore, the most susceptible to leverage in the form of iteration.

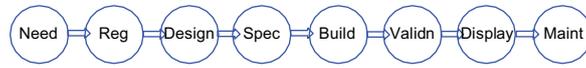

Figure 2.5 Phases in the Waterfall Model (Goguen, 1994).

Goguen (1994) complains that alternative process models still assume a division into phases and entirely ignore the characteristics of situatedness, especially around the fact that code must often be delivered before completing requirements and a high-level design is frequently required in defining requirements. In this light, it is useful to compare Michael Porter's (1985) model of the value chain with Stephen Kline's (1989) chain-link mode. While both authors divide the process into distinct phases (inbound logistics, operations, outbound logistics, marketing, and sales and services for Porter, and market finding and perception of needs, synthetic design, detailed design and test, redesign and produce, distribution, and market for Kline), Porter considers where value-added is contributed during production and service delivery, whereas Kline looks at how research contributes to the design and development of new products. Kline's mode has feedback loops connecting each succeeding phase to its predecessor. However, the thickest feedback loop extends from distribution and marketing back to market finding.

If we consider the thickest of the feedback loops depicted in Kline's "chain link" model of innovation, it would seem to mirror Goguen's (1994) notion of "requirements engineering." The framework of concurrency, communication and interaction that has been articulated above in the discussion of computational calculi and process algebras, plays an essential role in supporting these aspects of new product development within the UCS. The very same calculi assist in the management of both operational activity and innovation, irrespective of whether a particular firm has adopted integrated design or agile, lean, or iterative forms of new product development. Once again fundamental notions of concurrency (formally embodied in the concept of bisimulation) come to the fore when allocating resources and accounting for trade-offs between innovation-related and operational activities. Through their enabling of real-time simulation, communication and interaction, a raft of semantic technologies and various forms of cognitive computing can contribute to new product development in different ways, especially by nurturing new forms of co-creation between producers and users and by providing new forms of business intelligence that has been extracted from the WWW.

From a broader public policy perspective, Mazzucato and her collaborators (2015) have warned that the contemporary phenomenon of "financialization," defined both in the US and on a global scale by a growing share of profit and value-added accounted for by the financial sector, has a serious downside. It has encouraged increasingly speculative and myopic forms of investment which have supported trade in financial assets rather than production. Accordingly, she claims that State Investment Banks such as the IBRD, KfW, Export Bank of Japan, BNDES, KDB, BDBC, China Development Bank have a crucial and compensatory role to play in compensating for this situation, which goes beyond the more conventional provision of countercyclical investment capital and development funds for infrastructure to encompass new venture support and what could be called a "challenge-role."

In this context, Mazzucato and Wray (2015) cite the mission-oriented finance provided by DARPA during the Eisenhower administration during the Cold War designed to "put a man on the moon" before this could be done by the Soviet Union (Mazzucato & Wray, 2015). An example of greater relevance to the concerns of this chapter could be State funding for environmental sustainability initiatives.

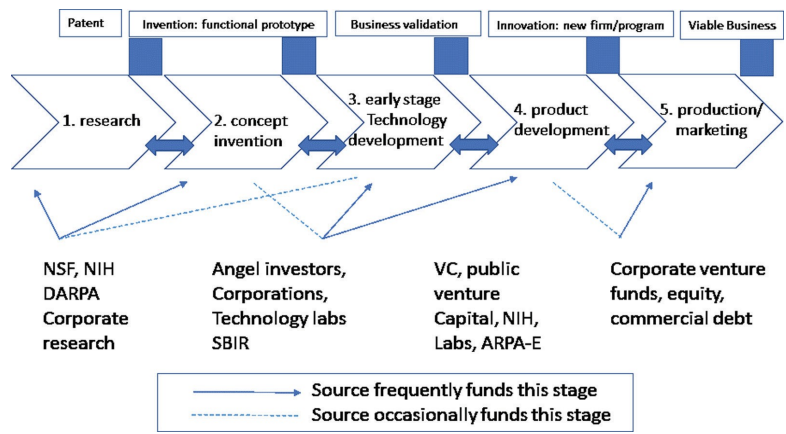

Figure 2.6 Public funding of innovation across the innovation chain (Mazzucato & Wray, 2015).

## 2.8 Organisational and Policy Implications

IoT and Big Data are fascinating developments and currently highly used keywords also in the context of Fog and Cloud Computing and Industry 4.0. The use of Big Data techniques for IoT comes natural as IoT produces large amounts of data. Similarly, as Deep Learning led to an unexpected performance jump in machine learning and pattern recognition, it is currently hard to predict where the combination of IoT and Big Data will lead in the future.

Once again, there is a need for a hardnosed and realistic appraisal of developments in the digital economy. The "dot-com" boom and slump occurred in an environment of extreme uncertainty over the respective upsides or downsides of the new technology. In much the same way, IoT now serves as a contemporary source of inflated expectations and asset price overvaluation.

Although much of what goes into effective STs is more integrative rather than path-breaking, the resulting constellation of carefully crafted search engines, databases, and diagrammatic reasoning modules could be a real source of gains in efficiency and effectiveness. And never before has the need for coordinated improvements in public policy been more compelling given (i) the scale of the transformations occurring within the digital economy, (ii) the complexity of problems we face around environmental sustainability and global warming, (iii) prospects for a partial retreat from the multilateral liberalization of trade, and (iv) impetus for a more nuanced approach to entrepreneurship in the public sector (e.g., compare Osborne & Gaebler, 1992, with Mazzucato & Wray, 2015).

Although deep learning on large data can now be conducted without the need for traditional "feature engineering," over-fitting remains a problem and next to parameter "tweaking," deep-learning systems still require careful planning, tailored implementations, and expert intervention. Human understanding of what is entailed by training draws both on non-mental intentional modes of comportment with the world and on pre-intentional encounters with the structural coherence of Being (as described by [Dreyfus, 2005]). For similar reasons, requirements engineering is the most obvious expression of how ontological and social uncertainties can only be resolved co-creatively in new product development through iterative and incremental forms of activity.

The system development literature reviewed above would suggest that our post-"waterfall" world of iterative and incremental New Product Development, characterized by zigzagging, overlapping phases, and feedback loops from end users to designers, is one that can increasingly be supported by a range of process algebras and calculi of interaction, communication and concurrency, along with a variety of techniques for diagrammatic reasoning and data visualization.

These insights suggest that a more critical stance should be adopted towards our promotion of the digital economy and our current obsessions with Big Data and the Internet of Things. Marketing strategists and computational pundits should never end up believing in their own promotional rhetoric. By the same token, managers and economic commentators who want to gain a deeper understanding of what has been outlined above should acknowledge that our old formal models of economic behavior at the level of the firm and the individual consumer, derived from 19th century energetics, need to be completely reconstituted from the ground up, based on more comprehensive and rigorous models of concurrency, communication, interaction, and open thermodynamic networks characterized by non-equilibrium steady-states, and decentralized control.

## Acknowledgments

The order of authors listed below is alphabetic. FA contributed to the engineering and computing aspects of this review and would like to acknowledge support through an Australian Government Research Training Program scholarship. SC addressed some aspects of machine learning and data analytics. JJ contributed theoretical and philosophical aspects.